\documentclass[aps,prd,showpacs,onecolumn,groupedaddress]{revtex4}
\usepackage{graphicx}
\usepackage{amsfonts}
\usepackage{amssymb}
\usepackage{amsmath}

\usepackage{ulem}
\usepackage{color}

\newcommand{\Slash}[1]{{\ooalign{\hfil/\hfil\crcr$#1$}}}

\begin{document}

\title{Relevant Gluonic Momentum for Confinement and \\
Gauge-Invariant Formalism with Dirac-mode Expansion}
\author{Hideo~Suganuma}
  \email{suganuma@ruby.scphys.kyoto-u.ac.jp}
  \affiliation{Department of Physics, Graduate School of Science,
  Kyoto University, \\
  Kitashirakawa-oiwake, Sakyo, Kyoto 606-8502, Japan}
\author{Shinya~Gongyo}
  \email{gongyo@ruby.scphys.kyoto-u.ac.jp}
  \affiliation{Department of Physics, Graduate School of Science,
  Kyoto University, \\
  Kitashirakawa-oiwake, Sakyo, Kyoto 606-8502, Japan}
\author{Takumi~Iritani}
  \email{iritani@ruby.scphys.kyoto-u.ac.jp}
  \affiliation{Department of Physics, Graduate School of Science,
  Kyoto University, \\
  Kitashirakawa-oiwake, Sakyo, Kyoto 606-8502, Japan}
\author{Arata Yamamoto}
  \affiliation{Department of Physics, The University of Tokyo, 
  Tokyo 113-0033, Japan}

\date{\today}
\begin{abstract}
We investigate the relevant gluon-momentum region 
for confinement in lattice QCD on $16^4$ 
at $\beta$=5.7, 5.8 and 6.0, based on the Fourier expansion. 
We find that the string tension $\sigma$, 
i.e., the confining force, is almost unchanged 
even after removing the high-momentum gluon component above 1.5GeV 
in the Landau gauge. 
In fact, the confinement property originates 
from the low-momentum gluon component below 1.5GeV, 
which is the upper limit to contribute to $\sigma$. 
In the relevant region, smaller gluon momentum component is 
more important for confinement. 
Next, we develop a manifestly gauge-covariant expansion 
of the QCD operator such as the Wilson loop, 
using the eigen-mode of the QCD Dirac operator 
$\Slash D=\gamma^\mu D^\mu$. 
With this method, we perform a direct analysis 
of the correlation between confinement and chiral symmetry breaking 
in lattice QCD on $6^4$ at $\beta$=5.6. 
As a remarkable fact, 
the confinement force is almost unchanged 
even after removing the low-lying Dirac modes, 
which are responsible to chiral symmetry breaking. 
This indicates that one-to-one correspondence does not 
hold for between confinement and chiral symmetry breaking 
in QCD.
\end{abstract}
%12.38.Aw --- General properties of QCD
%12.38.Gc --- Lattice QCD calculations
%14.70.Dj --- Gluons
\maketitle

\section{Introduction}

Nowadays, quantum chromodynamics (QCD) has been established 
as the fundamental gauge theory of the strong interaction. 
However, nonperturbative properties of low-energy QCD 
such as color confinement and chiral symmetry breaking \cite{NJL61}
are not yet well understood, which gives 
one of the most difficult problems in theoretical physics. 
The nonperturbative QCD has been studied 
in lattice QCD \cite{W74KS75,C7980,R05C11} 
and various analytical frameworks 
\cite{N74tH81,BC80,C8211,S94,SST95,T0709,ABP08}.

In this paper, using lattice QCD, 
we research for the origin of color confinement 
in terms of the relevant gluon-momentum component 
based on the Fourier expansion 
in the Landau gauge \cite{YS0809}.
We also investigate the correspondence between 
color confinement and chiral symmetry breaking 
using the Dirac-mode expansion in a gauge-invariant manner.

\section{Relevant Region of Gluonic Momentum for Color Confinement}

Many theoretical physicists consider that confinement phenomenon 
is brought by low-energy region of QCD, because of 
the strong QCD coupling in the infrared (IR) region. 
However, at the quantitative level, it is difficult to state 
the ``relevant energy region'' for confinement directly from QCD. 
Since nonperturbative phenomena are mainly brought by gluon dynamics, 
the key question here is 
{\it ``what is the relevant gluon-momentum region responsible 
for confinement?''}

In this section, to get the answer, 
we study quantitative lattice-QCD analysis for the relevant 
gluon-momentum region for color confinement \cite{YS0809}, 
based on the Fourier expansion of the link-variable.
Our method consists of the following five steps.

\vspace{0.4cm}

\noindent{\it 
Step 1. Generation of link-variable in the Landau gauge
}

\vspace{0.3cm}

We generate a gauge configuration on a $L^4$ lattice with 
the lattice spacing $a$ 
by the lattice-QCD Monte Carlo method 
under space-time periodic boundary conditions.
Here, we consider the link-variable $U_\mu(x)=e^{iagA_\mu(x)}$ 
fixed in the Landau gauge, where the fluctuation 
from gauge degrees of freedom is strongly suppressed, 
owing to the global suppression of gluon-field 
fluctuations \cite{IS0911}.

\vspace{0.4cm}

\noindent{\it
Step 2. Four-dimensional discrete Fourier transformation
}

\vspace{0.3cm}

By the discrete Fourier transformation, 
we define the ``momentum-space link-variable'', 
\begin{eqnarray}
{\tilde U}_{\mu}(p)=\frac{1}{N_{\rm site}}\sum_x 
U_{\mu}(x)\exp(i {\textstyle \sum_\nu} p_\nu x_\nu),
\end{eqnarray}
with the lattice-site number $N_{\rm site}$.
The momentum-space lattice spacing is given by
$a_p \equiv 2\pi/(La)$.

\vspace{0.4cm}

\noindent{\it
Step 3. Imposing a cut in the momentum space
}

\vspace{0.3cm}

We impose a cut on ${\tilde U}_{\mu}(p)$ 
in the momentum space, as shown in Fig.1. 
Outside the cut, we replace $\tilde U_{\mu}(p)$ 
by the free-field link-variable,
${\tilde U}^{\rm free}_{\mu}(p)=
\frac{1}{N_{\rm site}}\sum_x 
\exp(i {\textstyle \sum_\nu} p_\nu x_\nu)=\delta_{p0}$, 
corresponding to $U_\mu(x)=1$.
Then, the momentum-space link-variable 
${\tilde U}_{\mu}^{\Lambda}(p)$ 
with the cut is defined as 
\begin{equation}
\label{eq2}
{\tilde U}_{\mu}^{\Lambda}(p)= \Bigg\{
\begin{array}{cc}
{\tilde U}_{\mu}(p) & ({\rm inside \ cut})\\
{\tilde U}^{\rm free}_{\mu}(p)=\delta_{p0} & ~~({\rm outside \ cut}).
\end{array}
\end{equation}

\begin{figure}[h]
\begin{center}
\includegraphics[scale=0.4]{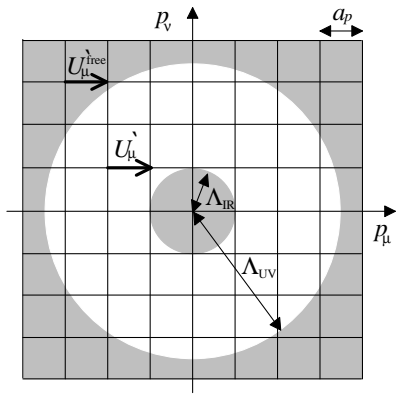}
\caption{
A schematic figure of the UV cut $\Lambda_{\rm UV}$ 
and the IR cut $\Lambda_{\rm IR}$ on momentum-space lattice, 
with the lattice spacing $a_p\equiv 2\pi/(La)$.
The momentum-space link-variable ${\tilde U}_{\mu}(p)$ is replaced 
by the free-field link-variable 
${\tilde U}^{\rm free}_{\mu}(p)=\delta_{p0}$ 
in the shaded cut regions.
}
\end{center}
\end{figure}

\vspace{0.4cm}

\noindent{\it
Step 4. Inverse Fourier transformation
}

\vspace{0.3cm}

To return to the coordinate space, 
we carry out the inverse Fourier transformation as
\begin{eqnarray}
U'_{\mu}(x)=\sum_p {\tilde U}_{\mu}^{\Lambda}(p)
\exp(-i {\textstyle \sum_\nu} p_\nu x_\nu).
\end{eqnarray}
Since this $U'_{\mu}(x)$ is not an SU(3) matrix, 
we project it onto an SU(3) element $U^{\Lambda}_{\mu}(x)$ by maximizing
$
{\rm ReTr}[U^{\Lambda}_{\mu}(x)^{\dagger}U'_{\mu}(x)].
$
Such a projection is often used in lattice QCD algorithms.
By this projection, we obtain the coordinate-space link-variable 
$U^{\Lambda}_{\mu}(x)$ with the cut, 
which is an SU(3) matrix and has the maximal overlap to $U'_{\mu}(x)$.

\vspace{0.4cm}

\noindent{\it
Step 5. Calculation of physical quantities
}

\vspace{0.3cm}

Using the cut link-variable $U^{\Lambda}_{\mu}(x)$, instead of $U_{\mu}(x)$, 
we calculate physical quantities as the expectation value
in the same way as original lattice QCD.

\vspace{0.55cm}

With this method in lattice-QCD framework, we quantitatively determine 
the relevant energy scale of color confinement, 
through the analyses of the $Q\bar Q$ potential.
The lattice QCD Monte Carlo simulations are performed on $16^4$ lattice 
at $\beta$=5.7, 5.8 and 6.0 at the quenched level \cite{YS0809}.

\begin{figure}[h]
\begin{center}
\includegraphics[scale=1]{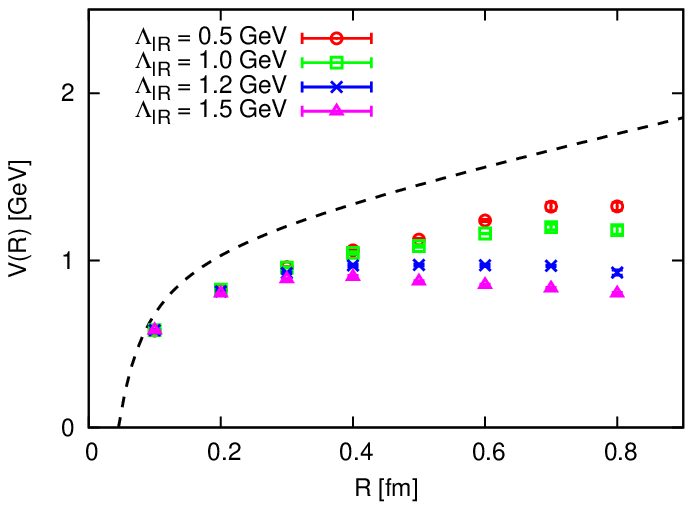}
\includegraphics[scale=1]{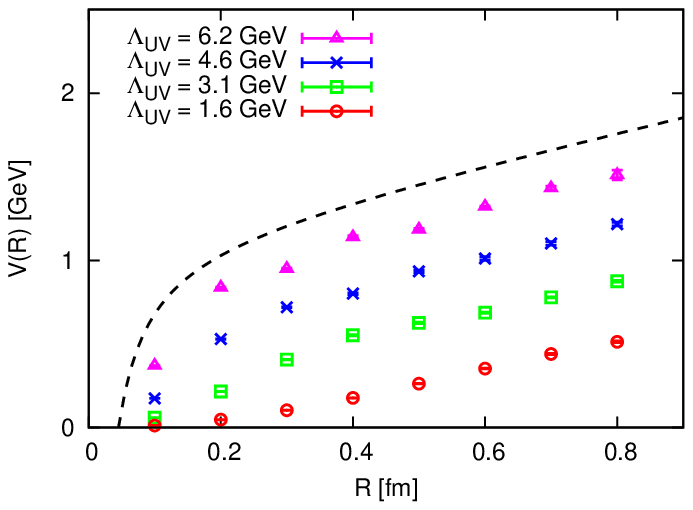}
\caption{
(a) The $Q\bar Q$ potential $V(R)$ with the IR cut 
$\Lambda_{\rm IR}$ plotted against the inter-quark distance $R$.
(b) The $Q\bar Q$ potential with the UV cut $\Lambda_{\rm UV}$.
Lattice QCD calculations are done on $16^4$ lattice 
with $\beta =6.0$, {\it i.e.}, $a\simeq 0.10$fm and 
$a_p \equiv 2\pi/(La) \simeq 0.77$GeV \cite{YS0809}.
The dashed line is the original $Q\bar Q$ potential in lattice QCD.
}
\end{center}
\end{figure}

Figure 2 (a) and (b) show the $Q\bar Q$ potential $V(R)$ 
with the IR cutoff $\Lambda_{\rm IR}$ 
and the UV cutoff $\Lambda_{\rm UV}$, respectively.
We get the following lattice-QCD results 
on the role of gluon momentum components.
\begin{itemize}
\item
By the IR cutoff $\Lambda_{\rm IR}$, as shown in Fig.2(a), 
the Coulomb potential seems to be unchanged, 
but the confinement potential is largely reduced \cite{YS0809}.
\item
By the UV cutoff $\Lambda_{\rm UV}$, as shown in Fig.2(b), 
the Coulomb potential is largely reduced, 
but the confinement potential is almost unchanged \cite{YS0809}.
\end{itemize}

\begin{figure}[h]
\begin{center}
\includegraphics[scale=1]{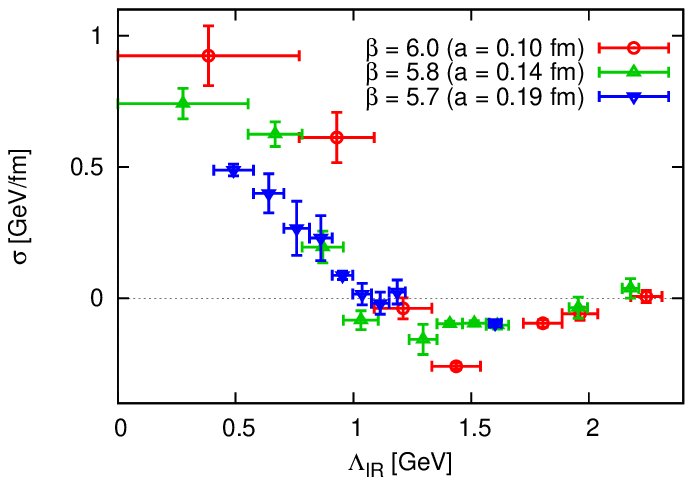}
\includegraphics[scale=1]{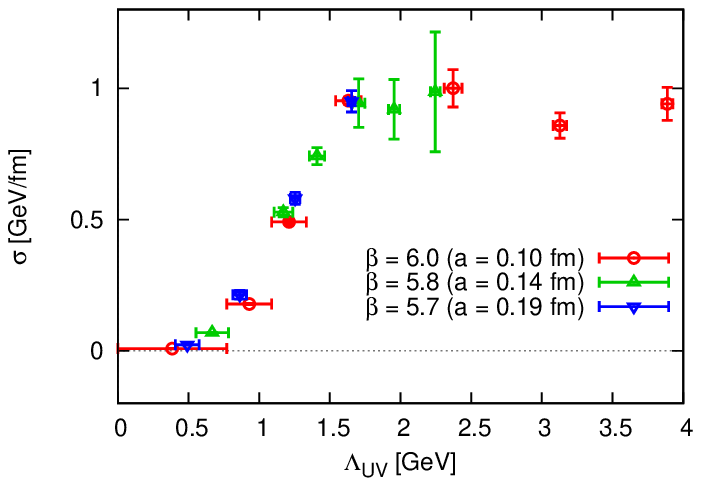}
\caption{
(a) The $\Lambda_{\rm IR}$-dependence and 
(b) the $\Lambda_{\rm UV}$-dependence of the string tension 
$\sigma$. 
The string tension is obtained from 
the asymptotic slope of the $Q\bar Q$ potential 
$V(R)$ with the IR/UV cutoff in lattice QCD on $16^4$ 
at $\beta$ =5.7, 5.8 and 6.0 \cite{YS0809}. 
The horizontal error-bar denotes the range from 
the discrete momentum, while 
the vertical error-bar is the statistical one.
The broken line denotes the original value of 
$\sigma \simeq 0.89$GeV/fm.
}
\end{center}
\end{figure}

Figure 3 shows the $\Lambda_{\rm IR}/\Lambda_{\rm UV}$-dependence 
of the string tension $\sigma$ obtained from the asymptotic slope 
of the $Q\bar Q$ potential $V(R)$ with the IR/UV cutoff.
Note that the ordinary QCD system without the cutoff corresponds to
$\Lambda_{\rm IR}=0$ and $\Lambda_{\rm UV}=+\infty$.
As shown in Fig.3(a), the string tension is significantly 
reduced by the IR-cutoff $\Lambda_{\rm IR}$ 
even for small values of $\Lambda_{\rm IR}$, 
and smaller gluon-momentum component seems to be 
more important for confinement. 

As a remarkable fact, the string tension is almost unchanged 
even after cutting off the high-momentum gluon component 
above 1.5GeV, as shown in Fig.3(b) \cite{YS0809}. 
In fact, the confinement property originates 
from the low-momentum gluon component below 1.5GeV, 
which is the upper limit to contribute to $\sigma$. 
Note here that the relevant region $|p|\le$ 1.5GeV 
for the confinement is only a small part of 
the total four-dimensional Brillouin zone of the gluon field, 
$-\pi/a < p_\mu \le \pi/a$ ($\mu$=1,2,3,4).
For example, at $\beta=6.0$ (i.e., $a \simeq 0.1{\rm fm}$), 
the relevant region $|p|\le$ 1.5GeV is 
less than 0.2\% in the total Brillouin zone, as shown in Fig.4.

\begin{figure}[h]
\begin{center}
\includegraphics[scale=0.5]{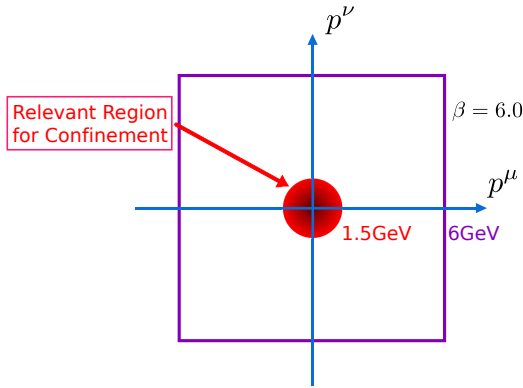}
\caption{
The relevant gluon-momentum region $|p|\le$ 1.5GeV for confinement 
in the Brillouin zone (BZ) 
$-\pi/a < p_\mu \le \pi/a(\simeq 6{\rm GeV})$ at $\beta=6.0$. 
The relevant region is less than 0.2\% in the four-dimensional BZ.
In the relevant region, smaller gluon-momentum component 
is more important for confinement.
}
\end{center}
\end{figure}

With the same method, we find also the relevant role of 
low-momentum gluons to chiral symmetry breaking 
in lattice QCD \cite{YS10}. However, the response pattern against the cut 
is rather different from that on confinement, 
and higher-momentum gluons also contribute to the chiral condensate.

\section{Gauge-Invariant Formalism with Dirac-mode Expansion:} 
\vspace{-0.45cm}
{\bf A Direct Investigation of Correlation between Confinement and 
Chiral Symmetry Breaking}

\vspace{0.4cm}

Next, we newly develop a manifestly gauge-covariant expansion 
of the QCD operator such as the Wilson loop, 
using the eigen-mode of the QCD Dirac operator 
$\Slash D=\gamma^\mu D^\mu$, and investigate 
the relation between confinement and chiral symmetry breaking. 

\vspace{-0.2cm}

\subsection{Gauge Covariant Expansion in QCD 
instead of Fourier Expansion}

The previous method is based on the Fourier expansion, i.e., 
the eigen-mode expansion of the momentum operator $p^\mu$. 
Because of the commutable nature of $[p^\mu, p^\nu]=0$, 
all the momentum $p^\mu$ can be simultaneously diagonalized. 
which is one of the strong merits of the Fourier expansion. 
Also it keeps Lorentz covariance. 

However, the Fourier expansion does {\it not} 
keep gauge invariance in gauge theories.
Therefore, for the use of the Fourier expansion in QCD, 
one has to select a suitable gauge such as the Landau gauge, 
where the gauge-field fluctuation is strongly suppressed 
in Euclidean QCD.

As a next challenge, we consider a gauge-invariant method, 
using a gauge-covariant expansion in QCD instead of the 
Fourier expansion. 
In fact, we consider a generalization of the Fourier expansion or 
an alternative expansion with keeping the gauge symmetry. 

A straight generalization is to use the covariant derivative operator 
$D^\mu$ instead of the derivative operator $\partial^\mu$. 
However, due to the non-commutable nature of $[D^\mu, D^\nu] \ne 0$, 
one cannot diagonalize all the covariant derivative 
$D^\mu$ ($\mu=1,2,3,4$) simultaneously, 
but only one of them can be diagonalized. 
For example, the eigen-mode expansion of 
$D_4$ keeps gauge covariance and is rather interesting, 
but this type of the expansion inevitably breaks the Lorentz covariance.
Then, we consider the eigen-mode expansion of 
the Dirac operator $\Slash D = \gamma^\mu D^\mu$ 
or $D^2 = D^\mu D^\mu$, since such an expansion keeps both 
gauge symmetry and Lorentz covariance.

In particular, the Dirac-mode expansion is rather interesting because 
the Dirac operator $\Slash D$ directly connects with 
chiral symmetry breaking via the Banks-Casher relation \cite{BC80} 
and its zero modes are related to the topological charge 
via the Atiyah-Singer index theorem \cite{S94}. 
Here, we mainly consider the manifestly gauge-invariant new method 
using the Dirac-mode expansion. 
Thus, the Dirac-mode expansion has some important merits.
\begin{itemize}
\item
The Dirac-mode expansion method manifestly keeps both 
gauge and Lorentz invariance.
\item
Each QCD phenomenon can be directly investigated in terms of 
chiral symmetry breaking.   
\end{itemize}

\subsection{Confinement and Chiral Symmetry Breaking:\\
Can we expect One-to-one Correspondence between them in QCD?}

In particular, it is rather interesting and important 
to examine the correlation between confinement 
and chiral symmetry breaking \cite{C8211,SST95,ABP08,M95W95,HFGHO08}, 
since the direct relation is not yet shown between them in QCD. 
The strong correlation between them has been suggested 
by the simultaneous phase transitions of deconfinement and 
chiral restoration in lattice QCD both 
at finite temperature \cite{R05C11} 
and in a small-volume box \cite{R05C11}.

The close relation between confinement and chiral symmetry breaking 
has been also suggested in terms of the monopole/vortex 
degrees of freedom \cite{SST95,M95W95,HFGHO08}, 
which topologically appears in QCD 
by taking the maximally Abelian/center gauge 
\cite{N74tH81,KSW87,SNW94}. 
For example, by removing the monopoles, 
confinement and chiral symmetry breaking are 
simultaneously lost in lattice QCD \cite{M95W95}, 
as schematically shown in Fig.5. 
This indicates an important role of the monopole 
to both confinement and chiral symmetry breaking, 
and these two nonperturbative QCD phenomena seem 
to be related via the monopole.

\begin{figure}[ht]
\begin{center}
\hspace{-0.5cm} \includegraphics[scale=0.6]{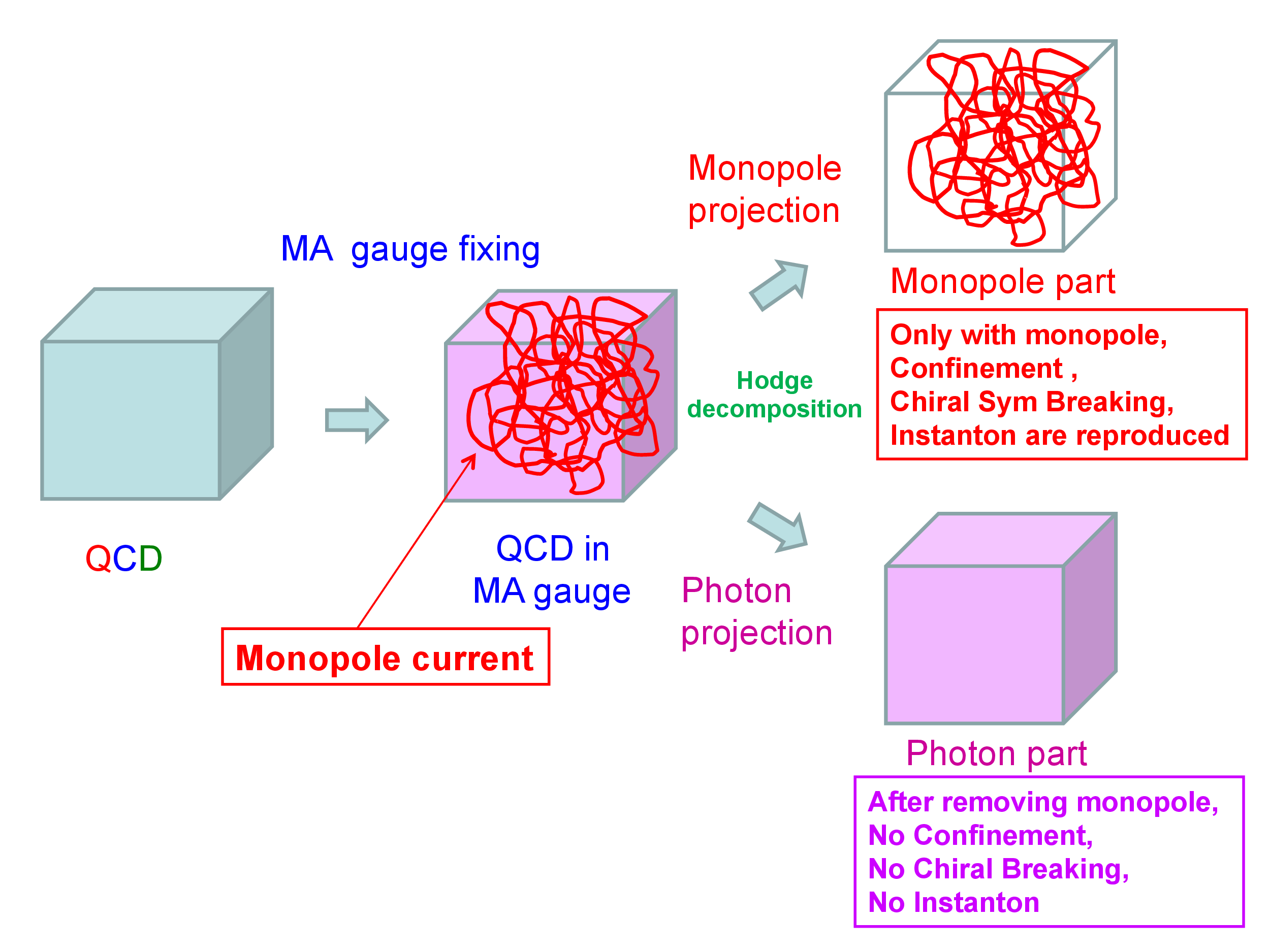}
\caption{
An illustration of 
the relevant role of monopoles to nonperturbative QCD. 
In the maximally Abelian gauge, 
QCD becomes Abelian-like due to the 
large off-diagonal gluon mass of about 1GeV \cite{AS99}, and 
there appears a global network of the monopole current \cite{KSW87,SNW94}. 
By the Hodge decomposition, the QCD system can be divided into 
the monopole part and the photon part. 
The monopole part has confinement \cite{SNW94},
chiral symmetry breaking \cite{M95W95} and instantons \cite{STSM95},
while the photon part does not have all of them.
}
\end{center}
\end{figure}

However, as a possibility, removing the monopoles 
may be ``too fatal'' for most nonperturbative properties. 
If this is the case, nonperturbative QCD 
phenomena are simultaneously lost by their cut. 

In fact, {\it if only the relevant ingredient of 
chiral symmetry breaking is carefully removed, 
how will be confinement?}
To get the answer, we perform a direct investigation between 
confinement and chiral symmetry breaking, 
using the Dirac-mode expansion.

\subsection{Eigen-mode of Dirac Operator in Lattice QCD}

In lattice QCD with spacing $a$, 
the Dirac operator 
$\Slash D = \gamma_\mu D_\mu$ is expressed with $U_\mu(x)$ as
\begin{eqnarray}
      \Slash{D}_{x,y} 
      \equiv \frac{1}{2a} \sum_{\mu=1}^4 \gamma_\mu 
\left[ U_\mu(x) \delta_{x+\hat{\mu},y}
        - U_{-\mu}(x) \delta_{x-\hat{\mu},y} \right],
\end{eqnarray}
where $U_{-\mu}(x)\equiv U^\dagger_\mu(x-\hat \mu)$.
In the use of hermite $\gamma$-matrix 
$\gamma_\mu^\dagger=\gamma_\mu$, 
$\Slash D$ is anti-hermite and satisfies 
$\Slash D_{y,x}^\dagger=-\Slash D_{x,y}$.
The normalized eigen-state $|n \rangle$ 
of the Dirac operator $\Slash D$ is introduced as 
\begin{eqnarray}
\Slash D |n\rangle =i\lambda_n |n \rangle
\end{eqnarray}
with $\lambda_n \in {\bf R}$.
Because of $\{\gamma_5,\Slash D\}=0$, the state 
$\gamma_5 |n\rangle$ is also an eigen-state of $\Slash D$ with the 
eigenvalue $-i\lambda_n$. 
The Dirac eigenfunction $\psi_n(x)\equiv\langle x|n \rangle$ 
obeys $\Slash D \psi_n(x)=i\lambda_n \psi_n(x)$, 
and its explicit form of the eigenvalue equation in lattice QCD is 
\begin{eqnarray}
\frac{1}{2a} \sum_{\mu=1}^4 \gamma_\mu
[U_\mu(x)\psi_n(x+\hat \mu)-U_{-\mu}(x)\psi_n(x-\hat \mu)]
=i\lambda_n \psi_n(x).
\end{eqnarray}
The Dirac eigenfunction $\psi_n(x)$ can be 
numerically obtained in lattice QCD, besides a phase factor. 

According to 
$U_\mu(x) \rightarrow V(x) U_\mu(x) V^\dagger (x+\hat\mu)$, 
the gauge transformation of $\psi_n(x)$ is found to be 
\begin{eqnarray}
\psi_n(x)\rightarrow V(x) \psi_n(x),
\label{eq:GTprop}
\end{eqnarray}
which is the same as that of the quark field.
To be strict, for the Dirac eigenfunction, 
there can appear an irrelevant $n$-dependent global phase factor 
as $e^{i\varphi_n[V]}$, 
according to the arbitrariness of the definition of $\psi_n(x)$.

Note that the quark condensate 
$\langle\bar qq \rangle$, the order parameter of 
chiral symmetry breaking, is given by 
the zero-eigenvalue density $\rho(0)$ 
of the Dirac operator, 
via the Banks-Casher relation \cite{BC80}, 
\begin{eqnarray}
\langle \bar qq \rangle=-\lim_{m \to 0} \lim_{V \to \infty} 
\pi\rho(0).
\end{eqnarray}
Here, 
$\rho(\lambda)\equiv 
\frac1V\sum_{n}\langle \delta(\lambda-\lambda_n)\rangle$ 
is the spectral density of the Dirac operator.
Also, the zero-mode number asymmetry of 
the Dirac operator $\Slash D$ is equal to 
the topological charge (the instanton number)
$Q \equiv \frac{g^2}{16\pi^2}\int d^4x 
\ {\rm Tr} \ (G_{\mu\nu} \tilde G_{\mu\nu})$, 
which is known as the Atiyah-Singer index theorem, 
Index($\Slash D$)=$Q$ \cite{S94}.

\subsection{Operator Formalism in Lattice QCD}

To keep the gauge symmetry, careful treatments are necessary, 
since naive approximations may break the gauge symmetry. 
Here, we take the ``operator formalism'', as explained below.

We define the link-variable operator $\hat U_\mu$ 
by the matrix element of 
\begin{eqnarray}
\langle x |\hat U_\mu|y\rangle =U_\mu(x)\delta_{x+\hat \mu,y}.
\end{eqnarray}
The Wilson-loop operator $\hat W$ is defined as the product of 
$\hat U_\mu$ along a rectangular loop,
\begin{eqnarray}
\hat W \equiv \prod_{k=1}^N \hat U_{\mu_k}
=\hat U_{\mu_1}\hat U_{\mu_2} \cdots \hat U_{\mu_N}.
\end{eqnarray}
For arbitrary loops, one finds $\sum_{k=1}^N \hat \mu_k=0$.
We note that the functional trace of the Wilson-loop operator 
$\hat W$ is proportional to the ordinary vacuum expectation value 
$\langle W \rangle$ of the Wilson loop:
\begin{eqnarray}
{\rm Tr} \ \hat W&=&{\rm tr}\sum_x \langle x |\hat W|x \rangle
={\rm tr}\sum_x \langle x| \hat U_{\mu_1}\hat U_{\mu_2} 
\cdots \hat U_{\mu_N}|x\rangle \nonumber\\
&=& {\rm tr} \sum_{x_1, x_2, \cdots, x_N }
\langle x_1| \hat U_{\mu_1}|x_2 \rangle
\langle x_2| \hat U_{\mu_2}|x_3 \rangle
\langle x_3| \hat U_{\mu_3}|x_4 \rangle
\cdots \langle x_N|\hat U_{\mu_N}|x_1\rangle \nonumber\\
&=&{\rm tr} \sum_x 
\langle x| \hat U_{\mu_1}|x+\hat \mu_1 \rangle
\langle x+\hat \mu_1| \hat U_{\mu_2}|x+\sum_{k=1}^2\hat \mu_k \rangle
\cdots \langle x+\sum_{k=1}^{N-1}\hat \mu_k|\hat U_{\mu_N}|x\rangle \nonumber\\
&=&\sum_x {\rm tr}\{ U_{\mu_1}(x) U_{\mu_2}(x+\hat \mu_1)
U_{\mu_3}(x+\sum_{k=1}^2 \hat \mu_k)
\cdots U_{\mu_N}(x+\sum_{k=1}^{N-1} \hat \mu_k)\} \nonumber\\
&=&\langle W \rangle \cdot {\rm Tr}\ 1.
\label{eq:TrWLO}
\end{eqnarray}
Here, ``Tr'' denotes the functional trace, 
and ``tr'' the trace over SU(3) color index.

The Dirac-mode matrix element of the link-variable operator 
$\hat U_{\mu}$ can be expressed with $\psi_n(x)$:
\begin{eqnarray}
\langle m|\hat U|n \rangle=\sum_x\langle m|x \rangle 
\langle x|\hat U_{\mu}|x+\hat \mu \rangle \langle x+\hat \mu|n\rangle
=\sum_x \psi_m^\dagger(x) U_\mu(x)\psi_n(x+\hat \mu).
\end{eqnarray}
Although the total number of the matrix element is very huge, 
the matrix element is calculable and gauge invariant, 
apart from an irrelevant phase factor.
Using the gauge transformation (\ref{eq:GTprop}), we find 
the gauge transformation of the matrix element as 
\begin{eqnarray}
\langle m|\hat U_\mu|n \rangle
&=&\sum_x \psi^\dagger_m(x)U_\mu(x)\psi_n(x+\hat\mu) \nonumber\\
&\rightarrow&
\sum_x\psi^\dagger_m(x)V^\dagger(x)\cdot V(x)U_\mu(x)V^\dagger(x+\hat \mu)
\cdot V(x+\hat \mu)\psi_n(x+\hat \mu) \nonumber\\
&=&\sum_x\psi_m^\dagger(x)U_\mu(x)\psi_n(x+\hat \mu)
=\langle m|\hat U_\mu|n\rangle.
\end{eqnarray}
To be strict, there appears an $n$-dependent global phase factor, 
corresponding to the arbitrariness of the phase in the basis 
$|n \rangle$. However, this phase factor cancels 
as $e^{-i\varphi_n} e^{i\varphi_n}=1$ 
between $|n \rangle$ and $\langle n |$, and does not appear 
for QCD physical quantities including the Wilson loop.

\subsection{Dirac-mode Expansion and Projection}

From the completeness of the Dirac-mode basis, 
$\sum_n|n\rangle \langle n|=1$, we get 
$\hat O=\sum_m\sum_n |m \rangle \langle m|\hat O|n \rangle \langle n|$ 
for arbitrary operators.
Based on this relation, the Dirac-mode expansion and projection 
can be defined. We define the projection operator $\hat P$ 
which restricts the Dirac-mode space, 
\begin{eqnarray}
\hat P\equiv \sum_{n \in A}|n\rangle \langle n|,
\end{eqnarray} 
where $A$ denotes arbitrary set of Dirac modes. 
In $\hat P$, the arbitrary phase cancels 
between $|n\rangle$ and $\langle n|$. 
One finds $\hat P^2=\hat P$ and $\hat P^\dagger =\hat P$.
The typical projections are 
IR-cut and UV-cut of the Dirac modes:
\begin{eqnarray}
\hat P_{\rm \ IR} \equiv 
\sum_{|\lambda_n| \ge \Lambda_{\rm IR}}|n \rangle \langle n|,
\qquad 
\hat P_{\rm \ UV} \equiv 
\sum_{|\lambda_n| \le \Lambda_{\rm UV}}|n \rangle \langle n|.
\end{eqnarray} 

Using the projection operator $\hat P$, we define 
the Dirac-mode projected link-variable operator, 
\begin{eqnarray}
\hat U^P_\mu \equiv \hat P \hat U_\mu \hat P
=\sum_{m \in A}\sum_{n \in A} 
|m\rangle \langle m|\hat U_\mu|n\rangle \langle n|.
\end{eqnarray}
During this projection, there appears some nonlocality in general, 
but it would not be important for the argument of 
large-distance properties such as confinement. 
From the Wilson-loop operator 
$\hat W \equiv \prod_{k=1}^N\hat U_{\mu_k}$, 
we define the Dirac-mode projected Wilson-loop operator,
\begin{eqnarray}
\hat W^P &\equiv& \prod_{k=1}^N \hat U^P_{\mu_k}
=\hat U^P_{\mu_1}\hat U^P_{\mu_2}\cdots \hat U^P_{\mu_N} 
=\hat P \hat U_{\mu_1} \hat P \hat U_{\mu_2} \hat P 
\cdots \hat P \hat U_{\mu_N} \hat P \nonumber\\
&=&\sum_{n_1, n_2, \cdots, n_{N+1} \in A} 
|n_1 \rangle \langle n_1| \hat U_{\mu_1}
|n_2 \rangle \langle n_2| \hat U_{\mu_2} |n_3 \rangle \cdots
\langle n_N| \hat U_{\mu_N}
|n_{N+1} \rangle \langle n_{N+1}|.
\end{eqnarray}
Then, we obtain the functional trace of 
the Dirac-mode projected Wilson-loop operator, 
\begin{eqnarray}
{\rm Tr} \ \hat W^P &=& {\rm Tr} \ \prod_{k=1}^N \hat U^P_{\mu_k}
={\rm Tr} \ \hat U^P_{\mu_1}\hat U^P_{\mu_2}\cdots \hat U^P_{\mu_N} 
={\rm Tr} \ \hat P \hat U_{\mu_1} \hat P \hat U_{\mu_2} \hat P 
\cdots \hat P \hat U_{\mu_N} \hat P \nonumber\\
&=&{\rm tr} \sum_{n_1, n_2, \cdots, n_N \in A} 
\langle n_1| \hat U_{\mu_1} |n_2 \rangle 
\langle n_2| \hat U_{\mu_2} |n_3 \rangle \cdots
\langle n_N| \hat U_{\mu_N}|n_{1} \rangle,
\end{eqnarray}
which is manifestly gauge invariant. 
Here, the arbitrary phase factor 
cancels between $|n_k \rangle$ and $\langle n_k|$. 
Its gauge invariance is also numerically checked 
in the lattice QCD Monte Carlo calculation.

From ${\rm Tr} \ \hat W^P(R,T)$ corresponding to 
the $R \times T$ rectangular loop, 
we define the Dirac-mode projected inter-quark potential, 
\begin{eqnarray}
V^P(R)\equiv -\lim_{T \to \infty} \frac{1}{T}
{\rm ln} \{{\rm Tr} \ \hat W^P(R,T)\},
\end{eqnarray}
which is also manifestly gauge-invariant.
In the unprojected case of $\hat P=1$, 
the ordinary inter-quark potential is obtained 
apart from an irrelevant constant,
\begin{eqnarray}
V(R)=-\lim_{T \to \infty} \frac{1}{T}
{\rm ln} \{{\rm Tr} \ \hat W(R,T)\}
= -\lim_{T \to \infty} \frac{1}{T}
{\rm ln} \langle W(R,T)\rangle + {\rm irrelevant} \ {\rm const.},
\end{eqnarray}
because of ${\rm Tr} \ \hat W=\langle W \rangle 
\cdot {\rm Tr}\ 1$, as was derived in Eq.(\ref{eq:TrWLO}).

\subsection{Analysis of Confinement in terms of Dirac Modes in QCD}

We consider various projection space $A$ in the Dirac-mode space, 
e.g., IR-cut or UV-cut of Dirac modes. 
With this Dirac-mode expansion and projection formalism, 
we calculate the Dirac-mode projected inter-quark potential $V^P(R)$ 
in a gauge-invariant manner. 
In particular, using IR-cut of the Dirac modes, 
we directly investigate the role of low-lying Dirac modes 
to confinement, since the low-lying modes are 
responsible to chiral symmetry breaking.

As a technical difficulty of this formalism, we have to deal with 
huge dimensional matrices and their products.  
Actually, the total matrix dimension of 
$\langle m|\hat U_\mu|n\rangle$ is (Dirac-mode number)$^2$. 
On the $L^4$ lattice, the Dirac-mode number is 
$L^4 \times N_c \times$  4, 
which can be reduced to be $L^4 \times N_c$, 
using the Kogut-Susskind technique \cite{W74KS75,R05C11}.
% and the chiral property of $\Slash D$. 
%
Even for the projected operators, where the Dirac-mode space is 
restricted, the matrix is generally still huge. 
At present, we use a small-size lattice 
in the actual lattice QCD calculation.

We use SU(3) lattice QCD at $\beta=5.6$ 
(i.e., $a\simeq 0.25{\rm fm}$) on $6^4$ at the quenched level.
We show in Fig.6(a) the spectral density $\rho(\lambda)$ of 
the QCD Dirac operator $\Slash D$. 
The chiral property of $\Slash D$ leads to 
$\rho(-\lambda)=\rho(\lambda)$.
Figure~6(b) is the IR-cut Dirac spectral density 
$\rho_{\rm IR}(\lambda)\equiv 
\rho(\lambda)\theta(|\lambda|-\Lambda_{\rm IR})$ 
with the IR-cutoff $\Lambda_{\rm IR}=0.5a^{-1}\simeq 0.4{\rm GeV}$.
By removing the low-lying Dirac modes, 
the chiral condensate is largely reduced as 
$\langle \bar qq\rangle_{\Lambda_{\rm IR}}/
\langle \bar qq\rangle \simeq 0.02$
around the physical region of $m \simeq 5{\rm MeV}$.

\begin{figure}[h]
\begin{center}
\includegraphics[scale=0.5]{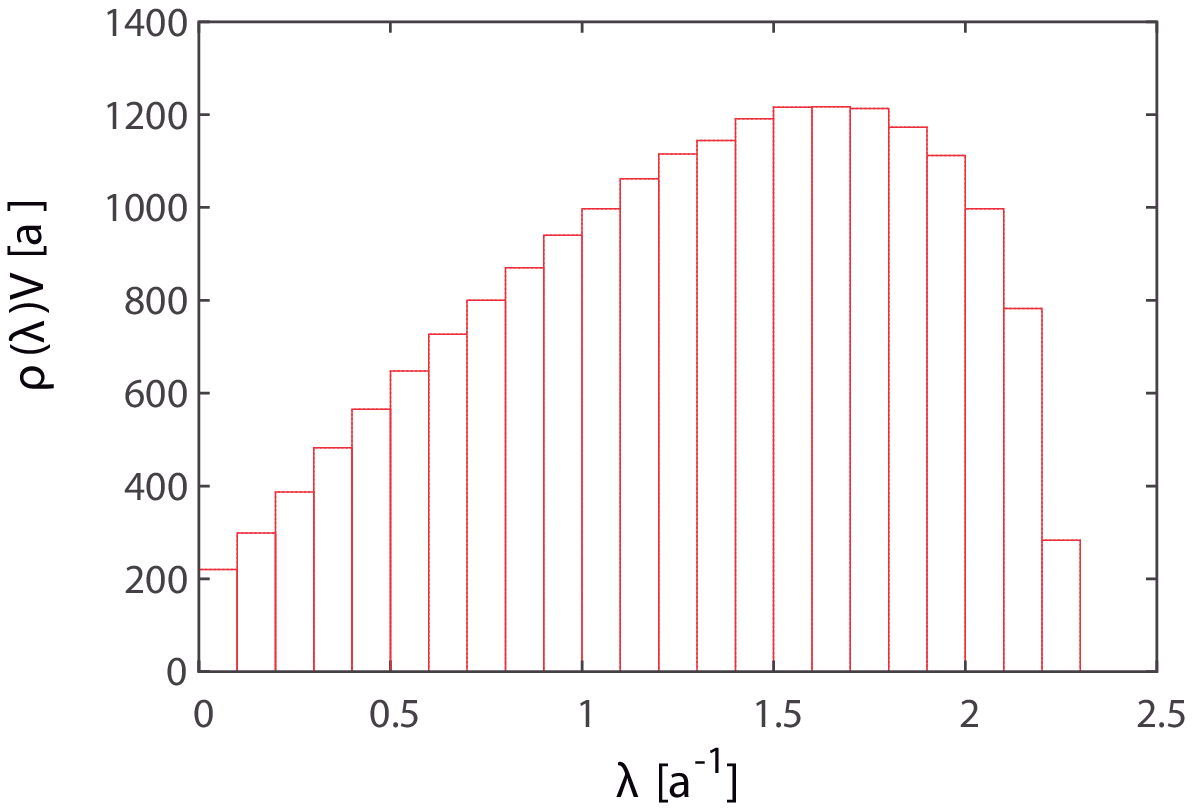}
\hspace{1cm}
\includegraphics[scale=0.5]{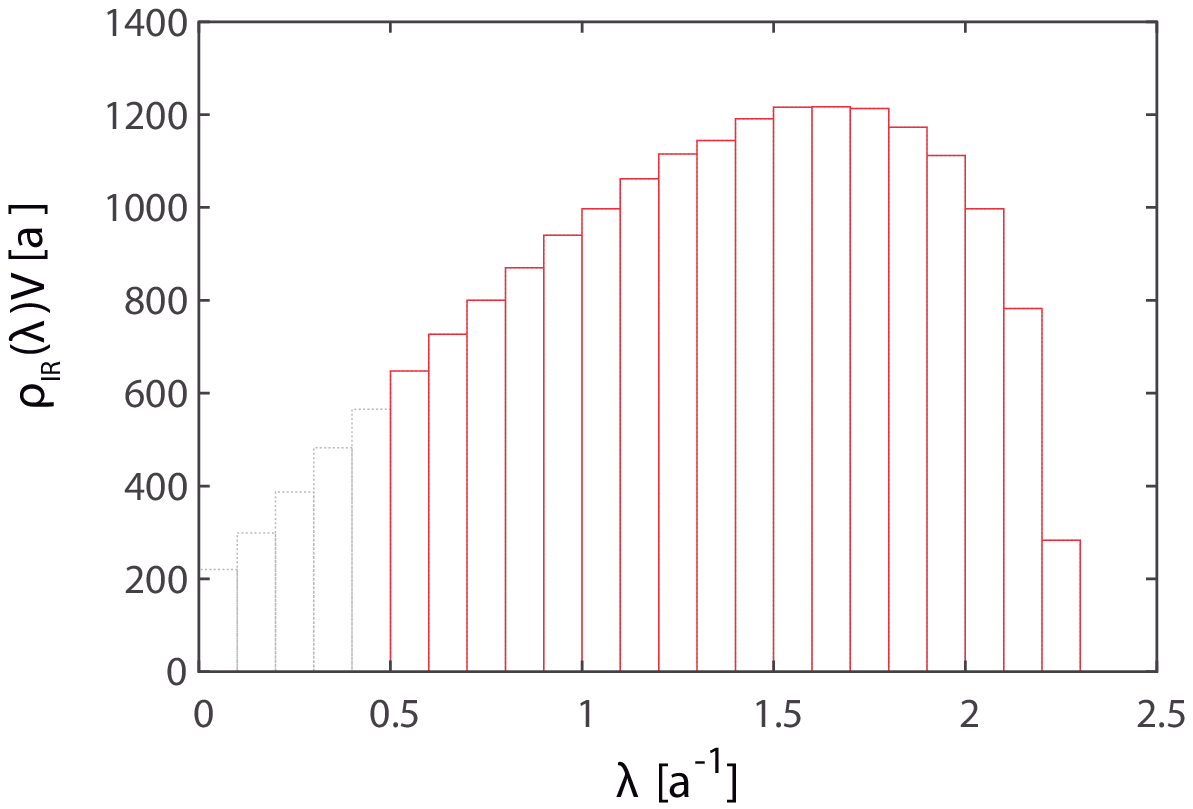}
\caption{
(a) The Dirac spectral density $\rho(\lambda)$ 
in lattice QCD at $\beta$=5.6 and $6^4$.
The volume $V$ is multiplied.
(b) The IR-cut Dirac spectral density 
$\rho_{\rm IR}(\lambda)\equiv 
\rho(\lambda)\theta(|\lambda|-\Lambda_{\rm IR})$ 
with the IR-cutoff $\Lambda_{\rm IR}=0.5a^{-1}\simeq 0.4{\rm GeV}$.
}
\end{center}
\end{figure}

Figure 7 shows the inter-quark potential $V^P(R)$ 
after removing low-lying Dirac modes, 
which is obtained in lattice QCD with the IR-cut of 
$\rho_{\rm IR}(\lambda)\equiv 
\rho(\lambda)\theta(|\lambda|-\Lambda_{\rm IR})$ 
with the IR-cutoff $\Lambda_{\rm IR}=0.5a^{-1}$.
Remarkably, no significant change is observed 
on the inter-quark potential besides an irrelevant constant, 
that is, the confining force is almost unchanged, 
even after removing the low-lying Dirac modes, 
which are responsible to chiral symmetry breaking.
This result indicates that one-to-one correspondence does not 
hold for confinement and chiral symmetry breaking in QCD.

\begin{figure}[h]
\begin{center}
\includegraphics[scale=0.7]{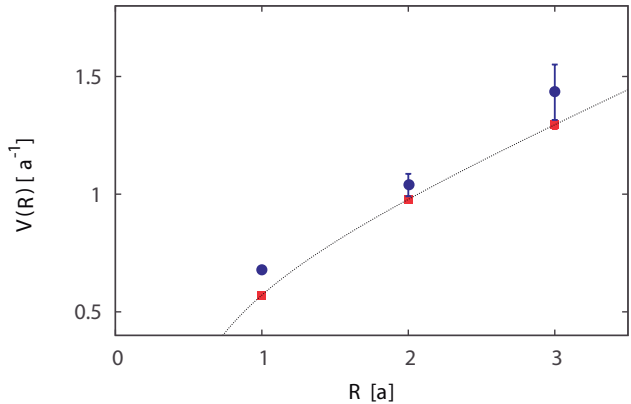}
\caption{
The circle symbol denotes 
the lattice QCD result of the inter-quark potential 
after removing low-lying Dirac modes, 
obtained with the IR-cut of 
$\rho_{\rm IR}(\lambda)\equiv 
\rho(\lambda)\theta(|\lambda|-\Lambda_{\rm IR})$ 
with the IR-cutoff $\Lambda_{\rm IR}=0.5a^{-1}$.
The square symbol denotes the original inter-quark potential.
The potential is almost unchanged 
even after removing the low-lying Dirac modes,
apart from an irrelevant constant.
}
\end{center}
\end{figure}

We also investigate the UV-cut of Dirac modes in lattice QCD, 
and find that the confining force is almost unchanged by the UV-cut. 
This result seems consistent with the pioneering lattice study 
of Synatschke-Wipf-Langfeld \cite{SWL08}: 
they found that the confinement potential is almost reproduced 
only with low-lying Dirac modes, 
using the spectral sum of the Polyakov loop \cite{G06BGH07}.
Furthermore, we examine ``intermediate-cut'', where 
a certain part of $\Lambda_1 < |\lambda_n| < \Lambda_2$ 
of Dirac modes is removed, 
and obtain almost the same confining force. 
Then, we conjecture that the ``seed'' of confinement is 
distributed not only in low-lying Dirac modes but also 
in a wider region of the Dirac-mode space. 

\subsection{Discussions on 
``Confinement $\ne$ Chiral Symmetry Breaking'' in QCD}

Here, we discuss the obtained result of 
``confinement $\ne$ chiral symmetry breaking''. 
As for their close relation via monopoles discussed in Sec.3.2, 
the monopole would be essential degrees of freedom for 
most nonperturbative QCD: confinement \cite{SNW94}, 
chiral symmetry breaking \cite{M95W95}, and instantons \cite{STSM95}.
In fact, removing the monopole would be ``too fatal''
for the nonperturbative properties, so that 
nonperturbative QCD phenomena are simultaneously lost by their cut. 

As shown in Fig.8, the Dirac zero-mode associated with an instanton 
is localized around it \cite{S94}. 
However, the localized objects are hard to contribute to 
the large-distance phenomenon of confinement, 
although such low-lying Dirac modes contribute to 
chiral symmetry breaking.

Recall that instantons contribute to chiral symmetry breaking,
but do not directly lead to confinement \cite{S94}. 
Then, as a thought experiment, 
if only instantons can be carefully removed from the QCD vacuum, 
confinement properties would be almost unchanged, 
but the chiral condensate is largely reduced, 
and accordingly some low-lying Dirac modes disappear. 
Thus, in this case, confinement is almost unchanged, 
in spite of the large reduction of low-lying Dirac modes.

\begin{figure}[h]
\begin{center}
\includegraphics[scale=0.4]{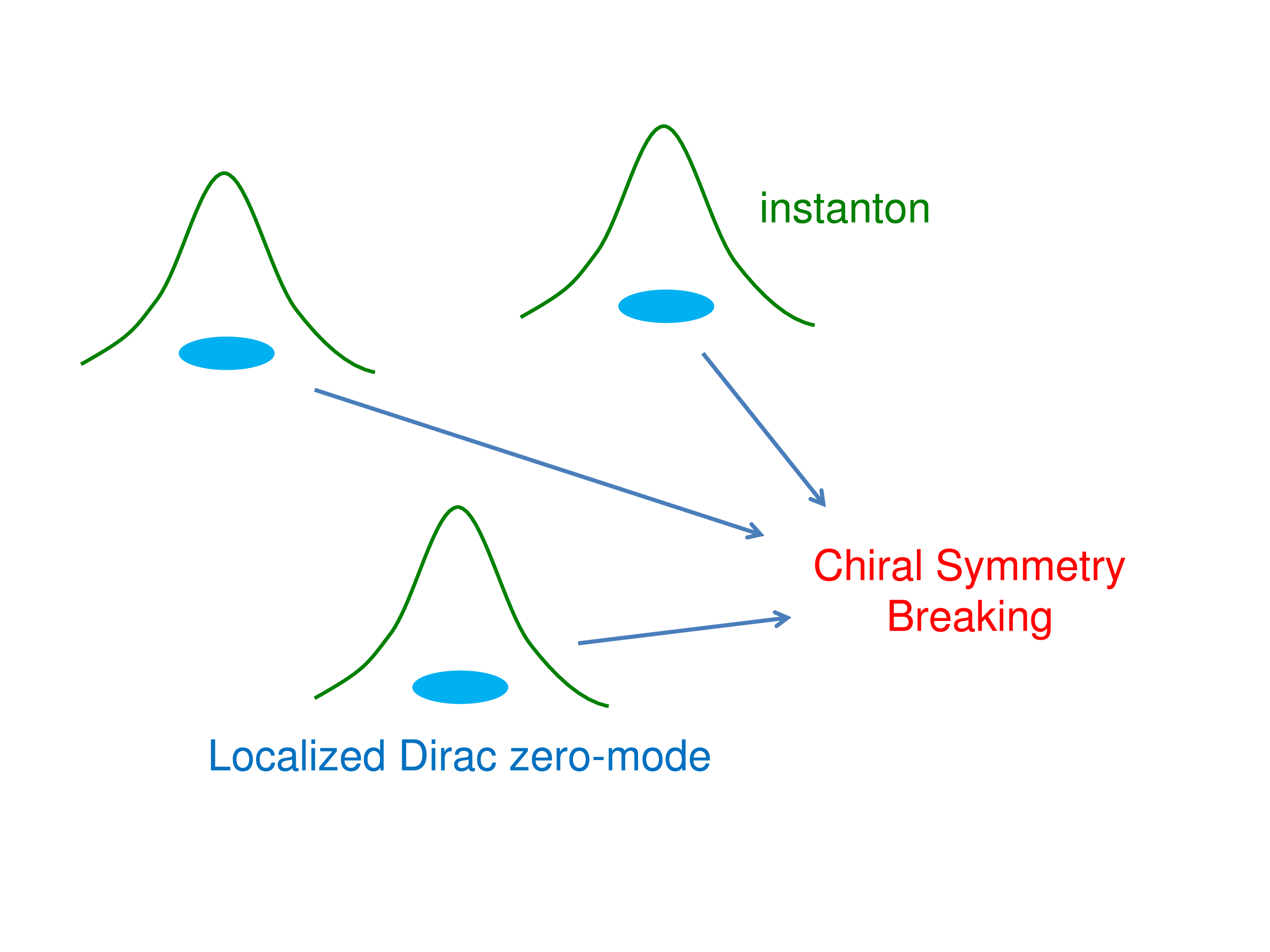}
\vspace{-1cm}
\caption{
Around each instanton, the Dirac zero-mode is localized, 
and such low-lying Dirac modes contribute to chiral symmetry breaking.
However, the localized objects are hard to contribute to 
confinement.
}
\end{center}
\end{figure}

If their relation is not one-to-one, 
richer phase structure is expected in QCD. 
For example, the phase transition point can be different 
between deconfinement and chiral restoration 
in the presence of strong electro-magnetic fields, 
because of their nontrivial effect on chiral symmetry \cite{ST9193}.

\section{Summary and Concluding Remarks}

First, we have studied the relevant gluon-momentum region 
for confinement in lattice QCD, 
based on the Fourier expansion. 
Remarkably, the string tension $\sigma$ is almost unchanged 
even after removing the high-momentum gluon component above 1.5GeV 
in the Landau gauge. We then have concluded that 
confinement originates from the low-momentum gluon component 
below 1.5GeV, which is the upper limit to contribute to $\sigma$. 

Second, we have developed a manifestly gauge-covariant expansion 
using the eigen-mode of the QCD Dirac operator 
$\Slash D=\gamma^\mu D^\mu$. 
With this method, we have performed a direct investigation 
of correspondence between 
confinement and chiral symmetry breaking in lattice QCD. 
As a remarkable fact, the confinement force is almost unchanged 
even after removing the low-lying Dirac modes, 
which are responsible to chiral symmetry breaking. 
This indicates that one-to-one correspondence does not 
hold for between confinement and chiral symmetry breaking in QCD.

\section*{Acknowledgements}
H.S. thanks Profs. J.M.~Cornwall, 
M.~Creutz, J.~Greensite, K.~Langfeld, 
J.~Papavassiliou, and E.T.~Tomboulis 
for useful discussions and suggestions. 
He thanks the organizers of QCD-TNT II. 
H.S. is supported in part by the Grant for Scientific Research 
[(C) No.~23540306, Priority Arias ``New Hadrons'' (E01:21105006)] 
from the Ministry of Education, Culture, Science and Technology 
(MEXT) of Japan.
The lattice QCD calculations are done on NEC SX-8R at Osaka University.

\end{document}